\documentstyle[twoside,fleqn,espcrc2]{article}

\newcommand{\AmS}{{\protect\the\textfont2
  A\kern-.1667em\lower.5ex\hbox{M}\kern-.125emS}}
\title{Determination of the asymptotic behaviour of the heavy flavour 
        coefficient functions in deep inelastic scattering}
\author{
        W.L. van Neerven, 
        \address{Instituut Lorentz, University of Leiden,
         PO Box, 9506, 2300 RA Leiden,
         The Netherlands.}%
         \thanks{Partially supported by EU contract CHRX-CT-92-0004.}
        M. Buza,
         \address{NIKHEF/UVA,POB 41882, NL-1009 DB Amsterdam
        P.O. Box 103, 1000 AC Amsterdam, The Netherlands.}%
        \thanks{Supported by the Foundation for Fundamental 
        Research on Matter (FOM).}%
        Y. Matiounine, J. Smith
        \address{Institute for Theoretical Physics, 
        State University of New York at Stony Brook,
        New York 11794-3840, USA.}%
        \thanks{Partially supported under the contract NSF-09888.}%
        and
         R. Migneron 
        \address{Department of Applied Mathematics,
         University of Western Ontario,
         London, Ontario, N6A 5B9, Canada.}%
         \thanks{Partially supported by the Netherlands organization 
         for scientific research (NWO) and by the National Sciences 
        and Engineering Research Council of Canada (NSERC).}%
           }
\begin{document}
\begin{abstract}
Using renormalization group techniques we have derived analytic
formulae for the next-to-leading order heavy-quark coefficient functions in
deep inelastic lepton hadron scattering. These formulae are only valid 
in the kinematic regime $Q^2 \gg m^2$, where $Q^2$ and $m^2$ stand for
the masses squared of the virtual photon and heavy quark respectively.
Some of the applications of these asymptotic formulae will be discussed.
\end{abstract}

\maketitle

\section{INTRODUCTION}
Deep inelastic electroproduction of heavy flavours is given by
\begin{equation}
e^-(\ell_1)+P(p)\rightarrow e^-(\ell_2)+Q(p_1)(\ \overline{Q}(p_2)\ )
+'X' \,.
\end{equation}
When the virtuality $-q^2 = Q^2 >0 \ (\ q=\ell_1-\ell_2\ )$ 
of the exchanged vector bosons is not too large $(\ Q^2\ll M_Z^2\ )$ the 
above reaction only gets a contribution 
from the virtual photon and we can neglect any weak effects
caused by the exchange of the Z-boson.  If the process is inclusive
with respect to the hadronic state $'X'$ as well as the heavy flavours
$Q(\bar Q)$, the unpolarized cross section is given by
\begin{eqnarray}
\frac{d^2\sigma}{dx dy}&=&\frac{2\pi\alpha^2}{(Q^2)^2}S\Big[\{1+(1-y)^2\}
F_2(x,Q^2,m^2)
\nonumber \\ && 
-y^2 F_L(x,Q^2,m^2)\Big] \,,
\end{eqnarray}
where $S$ denotes the square of the c.m. energy of the electron proton
system and the variables $x$ and $y$ are defined as
\begin{equation}
x=\frac{Q^2}{2p.q}\, (0<x\le 1)\, ,\,  y=\frac{p.q}{p.\ell_1}
\quad (0<y<1) \,,
\end{equation}
with 
\begin{equation}
-q^2=Q^2=xyS  ~.
\end{equation}
In the QCD improved parton model the heavy flavour contribution
to the hadronic structure functions, denoted by $F_i(x,Q^2,m^2)$
$(i=2,L)$, where $m$ stands for the heavy quark mass,
can be expressed as integrals over the partonic scaling 
variable. This yields the following results 
\begin{eqnarray}
&F_i(x,Q^2,m^2)&=x\int^{z_{max}}_x \frac{dz}{z}\Big[\frac{1}{n_f}
\sum^{n_f}_{k=1}e_k^2 
\nonumber \\ && 
\times\Big\{\Sigma\Big(\frac{x}{z},\mu^2\Big)
L^{\rm S}_{i,q}\Big(z,\frac{Q^2}{m^2},\frac{m^2}{\mu^2}\Big)
\nonumber \\ && 
+G\Big(\frac{x}{z},\mu^2\Big) L_{i,g}\Big(z,\frac{Q^2}{m^2},
\frac{m^2}{\mu^2}\Big)\Big\}
\nonumber \\ && 
+\Delta\Big(\frac{x}{z},\mu^2\Big)
L^{\rm NS}_{i,q}\Big(z,\frac{Q^2}{m^2},\frac{m^2}{\mu^2}\Big)\Big]
\nonumber \\ && 
+x~e_H^2\int^{z_{max}}_{x}\frac{dz}{z}
\nonumber \\ && 
\times\Big\{\Sigma\Big(\frac{x}{z},\mu^2\Big)
H_{i,q}\Big(z,\frac{Q^2}{m^2},\frac{m^2}{\mu^2}\Big)
\nonumber \\ && 
+G\Big(\frac{x}{z},\mu^2\Big)H_{i,g}\Big(z,\frac{Q^2}{m^2},
\frac{m^2}{\mu^2}\Big)\Big\} \,,
\nonumber \\ && 
\end{eqnarray}
where $z= Q^2/(s+Q^2)$ and $s$ is the square of the 
photon-parton centre-of-mass energy $(s \ge 4m^2)$. 
Here the upper boundary of the integration is given by
$z_{max}={Q^2}/(4 m^2 + Q^2)$. 

The function $G(z,\mu^2)$ stands for the gluon 
density whereas the flavour singlet and flavour non-singlet
combinations of the quark densities are given by $\Sigma(z,\mu^2)$ and
$\Delta(z,\mu^2)$ respectively.
In the above expressions the charges of the light quark and the heavy quark
are denoted by $e_i$ and $e_H$ respectively. Furthermore, $n_f$ stands
for the number of light quarks and $\mu$ denotes the mass factorization
scale, which we choose to be equal to the  renormalization scale.
The latter shows up in the running coupling constant defined by
$\alpha_s(\mu^2)$. The heavy quark coefficient
functions are denoted by $L_{i,j}$ and $H_{i,j}$ $(i=2,L$; $j=q,g)$.
The distinction between them can be traced back to the different photon-parton 
production processes from which they originate. The functions
$L_{i,j}$ are attributed to the reactions where the virtual photon
couples to the light quark, whereas the $H_{i,j}$ originate from the reactions
where the virtual photon couples to the heavy quark. 
Hence $L_{i,j}$ and $H_{i,j}$ in (5) are multiplied by $e_i^2$ and $e_Q^2$
respectively. The superscripts NS and S on the heavy quark coefficient 
functions refer to flavour non-singlet and flavour singlet respectively.
Furthermore the singlet quark coefficient functions $L^{\rm S}_{i,q}$ and
$H^{\rm S}_{i,q}$ can be split into non-singlet and purely singlet (PS)
parts,i.e.,
\begin{eqnarray}
L^{\rm S}_{i,q} = L^{\rm NS}_{i,q} + L^{\rm PS}_{i,q} \,,
\end{eqnarray}
\begin{eqnarray}
H^{\rm S}_{i,q} = H^{\rm NS}_{i,q} + H^{\rm PS}_{i,q} \,,
\end{eqnarray}
with $H^{\rm NS}_{i,q}=0$ in all orders of perturbation theory.

In \cite{lrsn1} the heavy quark coefficient functions $L_{i,j}$ and $H_{i,j}$
have been exactly calculated up to $\alpha_s^2$. 
Expanding them in a power series in $(\alpha_s/4\pi)^k$ they receive
contributions from the following parton subprocesses
\begin{equation}
\gamma^*(q) + g(k_1) \rightarrow Q(p_1) + \overline{Q}(p_2) \,,
\end{equation}
\begin{equation}
\gamma^*(q) + g(k_1) \rightarrow g(k_2) + Q(p_1) + \overline{Q}(p_2) \,,
\end{equation}
and
\begin{equation}
\gamma^*(q) +q(\overline{q})(k_1)\rightarrow q(\overline{q})(k_2)
+ Q(p_1) + \overline{Q}(p_2) ~ .
\end{equation}
For reaction (9) one has to include the virtual gluon corrections
to the Born process (8). The contributions from (8) and (9)
to the heavy quark coefficient functions are denoted by $H^{(1)}_{i,g}$
and $H^{(2)}_{i,g}$ respectively. The parton subprocess (10) has two
different production mechanisms. The first one is given by the
Bethe-Heitler process (see figs. 5a,5b in \cite{lrsn1}) leading to
$H^{{\rm PS},(2)}_{i,q}$ and the second one can be attributed to the
Compton reaction (see figs. 5c,5d in \cite{lrsn1}). 
Notice that $L^{{\rm PS}}_{i,q}$ and $ L^{{\rm S}}_{i,g}$ are zero 
through order $\alpha_s^2$.  Then, from (6), one infers that 
$L^{{\rm NS},(2)}_{i,q} = L^{{\rm S},(2)}_{i,q}$. 
Finally we want to make the remark that there are no interference
terms between the Bethe-Heitler and the Compton reactions in (10)
if one integrates over all final state momenta.

The complexity of the second order heavy quark coefficient functions prohibits
publishing them in an analytic form, except for $L_{i,q}^{{\rm NS},(2)}$,
which is given in Appendix A of \cite{bmsmn}, so that they are only
available in large computer programs \cite{lrsn1}, involving two-dimensional
integrations. 
To shorten the long running time needed for the computation of
the structure functions in (5)
we have tabulated the coefficient functions in the form of a two 
dimensional array in the variables $\eta$ and $\xi$ in 
a different computer program \cite{rsn1}. These variables are
defined by 
\begin{equation}
\eta=\frac{(1-z)}{4z}\xi -1\qquad,\qquad \xi=\frac{Q^2}{m^2}  ~.
\end{equation}
However when $\xi \gg 1$ ($Q^2 \gg m^2$) numerical instabilities
appear so that it is desirable to have analytic expressions
for the heavy quark coefficient functions in that region.
Moreover it turns out that for $\xi > 10$ the asymptotic expressions
for $H^{(2)}_{2,g}$ (9) and $H^{{\rm PS},(2)}_{2,q}$ (10)
approach the exact ones so that the former can be used for charm production
at the HERA collidier provided $Q^2 > 22.5 $ $({\rm GeV}/c^2)$
($m_c = 1.5$ $({\rm GeV}/c)$.
Furthermore one can use these asymptotic formulae in the context of
the variable flavour number scheme as has been explained in
\cite{aot}.
\section{METHOD}
We will now explain how the asymptotic form $(Q^2 \gg m^2)$
of the heavy quark coefficient functions $H_{i,j}$ , $L_{i,j}$
(5) can be derived using the renormalization group and the operator product
expansion (OPE) techniques. Using these techniques one can avoid
the computation of the cumbersome Feynman integrals and phase
space integrals which arise in the calculation of the
processes in (8) - (10).
In the limit $Q^2 \gg m^2$ the heavy quark coefficient functions
behave logarithmically as
\begin{eqnarray}
H^{(k)}_{i,j} \Big( z, \frac{Q^2}{m^2}, \frac{m^2}{\mu^2}\Big) &=&
\sum_{l = 0}^k a^{(k),(l)}_{i,j}\Big(z,\frac{m^2}{\mu^2}\Big)
\nonumber \\ &&
\times\ln^l\frac{Q^2}{m^2}  ~,
\end{eqnarray}
with a similar expression for L$_{i,j}^{(k)} $.
The above large logarithms originate from collinear divergences which 
arise when $Q^2$ is kept fixed and $m^2 \rightarrow 0$. As has been shown
in \cite{bmsmn} each fixed order term in expression (12) can be written as
\begin{equation}
H_{i,j}\Big( \frac{Q^2}{m^2}, \frac{m^2}{\mu^2}\Big)
= A_{kj}\Big( \frac{m^2}{\mu^2}\Big) \otimes C_{i,k}
\Big(\frac{Q^2}{\mu^2}\Big)\,,
\end{equation}
where the power of $\alpha_s$ has to match on the left and right hand
sides. There is a similar expression for $L_{i,j}$ ($i=2,L$; $k,j = q,g$).
Notice that we have suppressed the dependence on the scaling 
variable $z$ in (12) and the convolution symbol $\otimes$ is defined by
\begin{eqnarray}
\Big(f\otimes g\Big)(z)&=&\int_0^1 dz_1\int_0^1 dz_2  ~
\delta(z-z_1z_2)
\nonumber \\ &&
\times f(z_1)g(z_2)  ~.
\end{eqnarray}

The light quark and gluon coefficient functions $C_{i,k}$ have been
calculated up to order $\alpha_s^2$ in \cite{zn}. The operator 
matrix elements (OME's) $A_{kj}$ are now also known up to the same order in
perturbation theory (see \cite{bmsmn}). They are defined by
\begin{equation}
A_{kj}\Big(\frac{m^2}{\mu^2}\Big) = <j|O_k|j> \,,
\end{equation}
where $O_k$ represent the local operators which show up in the operator
product expansion of the two electromagnetic currents which appear 
in the calculation of the process (1).
Notice that the OME's in (15) are finite which means that all
renormalizations and mass factorizations have already been carried out.
The last operation is needed because of the collinear divergences
which appear in the OME's when the external on-mass-shell massless
quarks and gluons are coupled to internal massless quanta.
The ultraviolet and collinear divergences are regulated by using the
method of $n$-dimensional regularization. The removal 
of these divergences has been done in the
$\overline{\rm MS}$-scheme. For the computation of the heavy quark 
coefficient functions $H^{(1)}_{i,g}$ (8) $H^{(2)}_{i,g}$ (9)
we need the OME's $A^{(1)}_{Qg}$ and $A^{(2)}_{Qg}$ respectively,
which are given by the Feynman graphs in figs.1,2 of \cite{bmsmn}.
The Bethe-Heitler coefficient functions 
$H^{{\rm PS},(2)}_{i,q}$ (10)
requires the calculation of $A^{{\rm PS},(2)}_{Qq}$
whereas for the Compton coefficient function $L^{{\rm NS},(2)}_{Qq}$ (10)
one has to compute $A^{{\rm NS},(2)}_{qq}$ 
The results for these OME's can be found in appendix C of \cite{bmsmn}.
Substitution of $A_{kj}$ and $C_{i,k}$ in (13) leads
to the asymptotic heavy quark coefficient functions which are presented in
Appendix D of \cite{bmsmn}.

\section{RESULTS}
We are now interested to find out at which values of $\xi$ (11) or
$Q^2$ the asymptotic coefficient functions approach the exact ones
computed in \cite{lrsn1} and \cite{rsn1}.
For that purpose we define the ratio $R^{(\ell)}_{i,j}$
which is given by
\begin{equation}
R^{(\ell)}_{i,j}\Big(z, \xi,\frac{m^2}{\mu^2}\Big)
= { 
{H^{{\rm exact},(\ell)}_{i,j} (z, \xi,m^2/\mu^2)}
\over
{H^{{\rm asymp},(\ell)}_{i,j} (z, \xi,m^2/\mu^2)} }\,,
\end{equation}
where $H^{\rm exact}_{i,j}$ and $H^{\rm asymp}_{i,j}$ stand for the exact
\cite{lrsn1}, \cite{rsn1} and asymptotic \cite{bmsmn} heavy
quark coefficient functions respectively.
Choosing $\mu^2 = m^2$ and the range $5 < \xi < 10^5$, we have plotted as
an example $R^{(2)}_{L,g}$ in fig.1 and $R^{(2)}_{2,g}$ in fig.2
for $z = 10^{-2}$ and $z=10^{-4}$. The reason that we have chosen these two
ratios is that the coefficient functions $H^{(2)}_{L,g}$ and
$H^{(2)}_{2,g}$ (9) constitute the 
\begin{figure}[tp]
\vspace{7cm}
\begin{picture}(7,7)
\includegraphics{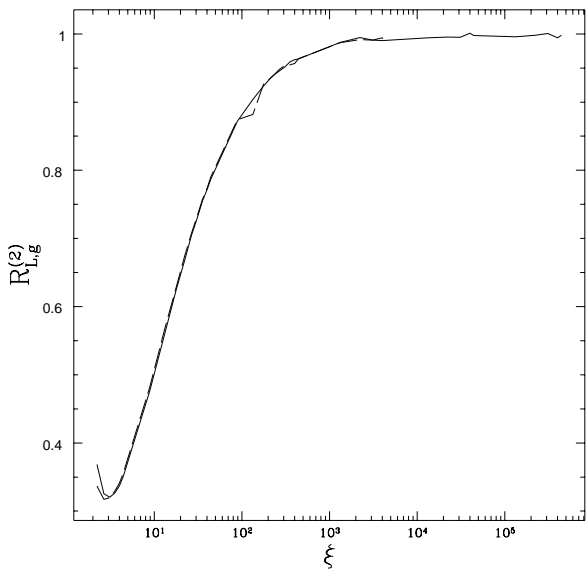}
\end{picture}
\caption{$R^{(2)}_{L,g}$ plotted as a function of $\xi$ for fixed
$z = 10^{-2}$ (solid line) and for $z = 10^{-4}$ (dashed line).}
\end{figure}
bulk of the radiative 
corrections to the Born reaction (8). From fig.1 we infer that 
$H^{{\rm exact},(2)}_{L,g}$ and
$H^{{\rm asymp},(2)}_{L,g}$ coincide when 
$\xi \ge \xi_{\rm min} = 10^3$ and there is essentially no difference between
the ratios for $z=10^{-2}$ and $z = 10^{-4}$. In the case of
$H^{{\rm exact},(2)}_{2,g}$ and
$H^{{\rm asymp},(2)}_{2,g}$  (see fig.2) the above $\xi$-value is much 
smaller and both coefficient functions coincide 
when $\xi \ge \xi_{\rm min}=10$,
which is quite insensitive to the values chosen for $z$. 
The reason why the convergence of 
$ R^{(2)}_{L,g}$ to one is so slow in comparison to
$ R^{(2)}_{2,g}$ is unclear. Apparently the logarithmic terms in
$H^{{\rm exact},(2)}_{L,g}$ start to dominate the coefficient 
functions at much larger values of $\xi$
than is the case for $H^{{\rm exact},(2)}_{2,g}$. 
A similar observation has been made for
$H^{{\rm PS},(2)}_{i,q}$. The small value found for $\xi_{\rm min}$ in
the case of $H_{2,g}$ is very interesting for charm production
where $F_2(x,Q^2,m_c^2)$ can be
measured with much higher accuracy than $F_L(x,Q^2,m^2)$.
Since $H^{(2)}_{2,g}$ dominates the radiative corrections to
$F_2(x,Q^2,m_c^2)$ one can state that for $Q^2 > 22.5$ $({\rm GeV}/c)^2$
$(m_c = 1.5$ ${\rm GeV}/c)$ the exact coefficient functions can be replaced
by their asymptotic ones. However before one can draw definite conclusions
about the dominance of the terms $ln^{l}(Q^2/m^2)$ on the level of the
structure functions ones first to convolute the heavy quark coefficient
functions with the parton densities ~(see (5) ). This will be done in future
work. If it turns out that the above logarithms also 
dominate $F_k(x,Q^2,m^2)$, in particular for $k=2$, then these terms 
have to be resummed using the renormalization group 
equations. This is done using the variable flavour
number scheme approach \cite{aot}. One of the features of this method is that
one has to define a charm density in the proton which is a convolution of the
OME's $A_{k,j}$ (15) and the light parton densities $\Sigma$ and $G$ in (5).
Hence for $Q^2>>m_c^2$ the charm quark behaves like a light parton provided
the large logarithmic terms dominate the proton structure functions in (5).
\begin{figure}[tp]
\vspace{7cm}
\begin{picture}(7,7)
\includegraphics{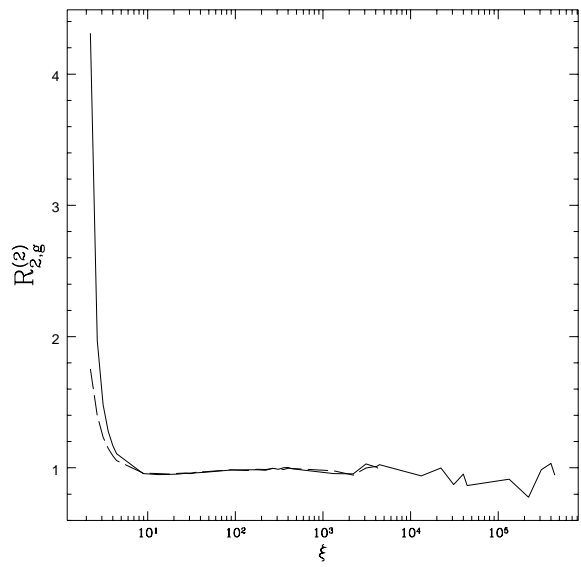}
\end{picture}
\caption{$R^{(2)}_{2,g}$ plotted as a function of $\xi$ for fixed
$z = 10^{-2}$ (solid line) and for $z = 10^{-4}$ (dashed line).}
\end{figure}
%
%

\end{document}